\documentclass[twocolumn,prl,tightenlines,superscriptaddress,showpacs]{revtex4-1}

\usepackage{amsmath}
\usepackage{amssymb,amsfonts,latexsym}
\usepackage{bm}
\usepackage[mathcal]{euscript}
\usepackage{graphicx}
\usepackage{epsfig}
\usepackage{color}
\usepackage{xfrac}
\usepackage{mathrsfs}

\usepackage{hyperref}
\usepackage{url}

\begin{document}

\title{Susceptibility of Orientationally-Ordered Active Matter to Chirality Disorder} 

\author{Bruno Ventejou}
\affiliation{Service de Physique de l'Etat Condens\'e, CEA, CNRS Universit\'e Paris-Saclay, CEA-Saclay, 91191 Gif-sur-Yvette, France}

\author{Hugues Chat\'{e}}
\affiliation{Service de Physique de l'Etat Condens\'e, CEA, CNRS Universit\'e Paris-Saclay, CEA-Saclay, 91191 Gif-sur-Yvette, France}
\affiliation{Computational Science Research Center, Beijing 100193, China}
\affiliation{Sorbonne Universit\'e, CNRS, Laboratoire de Physique Th\'eorique de la Mati\`ere Condens\'ee, 75005 Paris, France}

\author{Raul Montagne}
\affiliation{Departamento de Fisica, UFRPE, 52171-900 Recife, Pernambuco, Brazil}

\author{Xia-qing Shi}
\affiliation{Center for Soft Condensed Matter Physics and Interdisciplinary Research, Soochow University, Suzhou 215006, China}

\date{\today}

\begin{abstract}
We investigate the susceptibility of  
long-range ordered phases of two-dimensional dry aligning active matter to population disorder, 
taken in the form of a distribution of intrinsic individual chiralities.
Using a combination of particle-level models and hydrodynamic theories derived from them,
we show that while in finite systems all ordered phases resist a finite amount of such chirality disorder,
the homogeneous ones (polar flocks and active nematics) are unstable to any amount
of disorder in the infinite-size limit.
On the other hand, we find that the inhomogeneous solutions of the 
coexistence phase (bands) may resist a finite amount of chirality disorder even asymptotically. 
\end{abstract}

\maketitle

Many if not most active matter systems are made of interacting units that convert energy gathered
from their environment to displace themselves. 
In most models and theories, these self-propelled particles are taken identical and they evolve in a homogeneous medium.
Of course, particles in real systems are never strictly identical, 
nor do they move, in the common case where they are in direct contact with a substrate, in a pristine space.
To what extent the spectacular collective phenomena of active matter uncovered in models and theories
resist population or environment disorder is thus a valid question.

Recent works have made significant progress regarding the effects of heterogeneous substrates 
(spatial quenched disorder) on active matter
\cite{chepizhko2013optimal,reichhardt2014active,chepizhko2015active,quint2015topologically,morin2016distortion,pince2016disorder-mediated,sandor2017dynamic,morin2017diffusion,martinez2018collective,reichhardt2018clogging,toner2018swarming,toner2018hydrodynamic,das2018polar,das2018ordering,dor2019ramifications,ro2021disorder-induced,chardac2021emergence,duan2021breakdown}.
As often, attention focussed mostly on `polar flocks', the nickname for the homogeneous collective motion phase
exhibited by self-propelled particles locally aligning their velocities against some noise, as in the Vicsek model and the Toner-Tu theory \cite{vicsek1995novel,toner1995long,toner1998flocks,toner2012reanalysis,marchetti2013hydrodynamics,chate2020dry}.
%%%
Even though the matter is not yet fully settled, it was found that the true long-range polar order 
present even in two-dimensions (2D) 
is deeply modified and sometimes broken by any amount of quenched disorder. 

Is orientationally-ordered active matter equally susceptible to population disorder? Only a few active systems with 
heterogeneous population have been studied so far
\cite{menzel2012collective,degond2014hydrodynamics,ariel2015order-disorder,copenhagen2016self-organized,yllanes2017how,levis2019activity,levis2019simultaneous,pattanayak2020speed,bera2020motile,moore2021chiral,fruchart2021non-reciprocal}. 
Again, most of these works deal with aligning self-propelled particles and investigate the fate of collective motion phases. 
The population disorder considered takes the form of either two subpopulations, 
each made of identical particles, or truly distributed disorder, with each particle assigned some individual parameter. 
While some of these works have revealed interesting phenomena, 
such as the non-reciprocal phase transitions of \cite{fruchart2021non-reciprocal}, others did consider the robustness of polar flocks
and all concluded, often implicitly, that they resist a finite amount of disorder. 
%%%
This conclusion was in particular reached for systems with chirality disorder, 
in which self-propelled particles each possess 
an intrinsic tendency to turn either clockwise (CW) or counterclockwise (CCW), 
but with the total population remaining globally achiral \cite{levis2019activity,levis2019simultaneous}. 

In short, polar flocks seem sensitive to spatial quenched disorder, 
but robust to finite amounts of population disorder. 
In the latter case, though, only relatively small systems were considered, and no finite-size study was
provided.

In this Letter, we thoroughly investigate the susceptibility of  
long-range ordered phases of 2D dry aligning active matter
to chirality disorder.
Using both particle level models and hydrodynamic theories derived from them,
we show that, asymptotically, any amount of this type of disorder breaks both polarly and nematically 
ordered homogeneous phases. On the other hand, we find that the traveling Vicsek bands characterizing the coexistence phase in the polar case may resist a finite amount of chirality disorder. 
We provide a brief description of the 
chirality-induced phases replacing the long-range ordered ones and rough phase diagrams, but defer their detailed study to a future publication \cite{TBP-KVM}.

%%%%%
We consider point particles $i=1,\ldots,N$, endowed with an intrinsic frequency $\omega_i$,
which move at constant speed $v_0$ in a square domain of linear size $L$
with periodic boundary conditions. Their positions ${\bf r}_i$ and orientations $\theta_i$ evolve in continuous time:
\begin{subequations}
\begin{align}
\dot{\bf r}_i &= v_0 \, {\bf e}(\theta_i) \\
\dot{\theta}_i &= \omega_i + \kappa \left\langle \sin\alpha(\theta_j - \theta_i) \right\rangle_{j\sim i} + \sqrt{2D_{\rm r}}\eta_i
\end{align}
\label{eq:kvm}
\end{subequations}
where ${\bf e}(\theta)$ is the unit vector along $\theta$, 
%%%%%%
the average $\langle\ldots\rangle_{j\sim i}$ is taken over all particles within unit distance of ${\bf r}_i$, 
and $\eta_i$ is a uniform white noise drawn in $[-\pi,\pi]$.
For $\alpha=1$, particles align ferromagnetically, like in the Vicsek model, while alignment is nematic for $\alpha=2$. 
Both cases are considered below. 
Finally, the individual frequencies $\omega_i$ are drawn from a zero-mean distribution. Here we study two cases: a Gaussian distribution of rms $\omega_0$ and a bimodal distribution where half of the particles have frequency $+\omega_0$ and the other half $-\omega_0$.

In the pure, disorderless case of identical particles ($\omega_i=0, \, \forall i$), one finds the typical phase diagram
of Vicsek-style models: at any global density $\rho_0=N/L^2$, decreasing the noise strength $D_{\rm r}$ or, equivalently, increasing the coupling strength $\kappa$, one transits from a disordered gas to a homogeneous  ordered liquid with non-trivial fluctuations. This transition is not direct, but via a coexistence phase in which particles organize themselves in dense, ordered bands evolving in a residual sparse gas \cite{chate2020dry}. 

%%%%%%%%%%%%%%%%%%%%%%%%%%%
\begin{figure}
\includegraphics[width=\columnwidth]{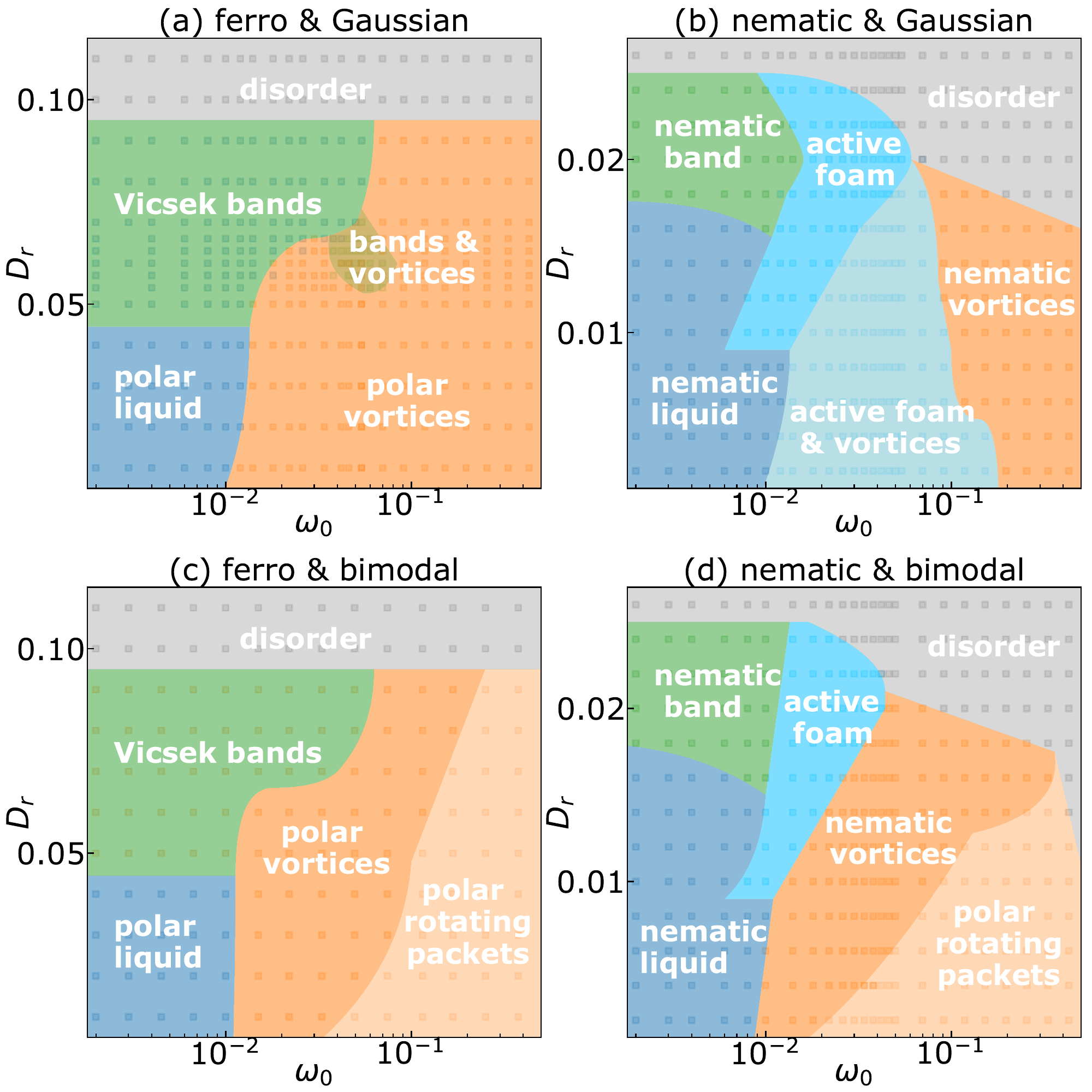}
	\caption{Phase diagrams in the $(\omega_0, D_{\rm r})$ plane ($\rho_0=\kappa=v_0=1$, $L=256$).
	(a,c) (left column) ferromagnetic alignment. 
	(b,d) (right column) nematic alignment. 
	Top row (a,b): Gaussian distribution; bottom row (c,d): bimodal distribution.
	(Details about the numerical protocol used to build these diagrams are given in \cite{SUPP}.)}
	\label{fig1}
\end{figure}
%%%%%%%%%%%%%%%%%%%%%%%%%%%

%%%%%%%%%%%%%%%%%%%%%%%%%%%
\begin{figure*}[t!]
\includegraphics[width=\textwidth]{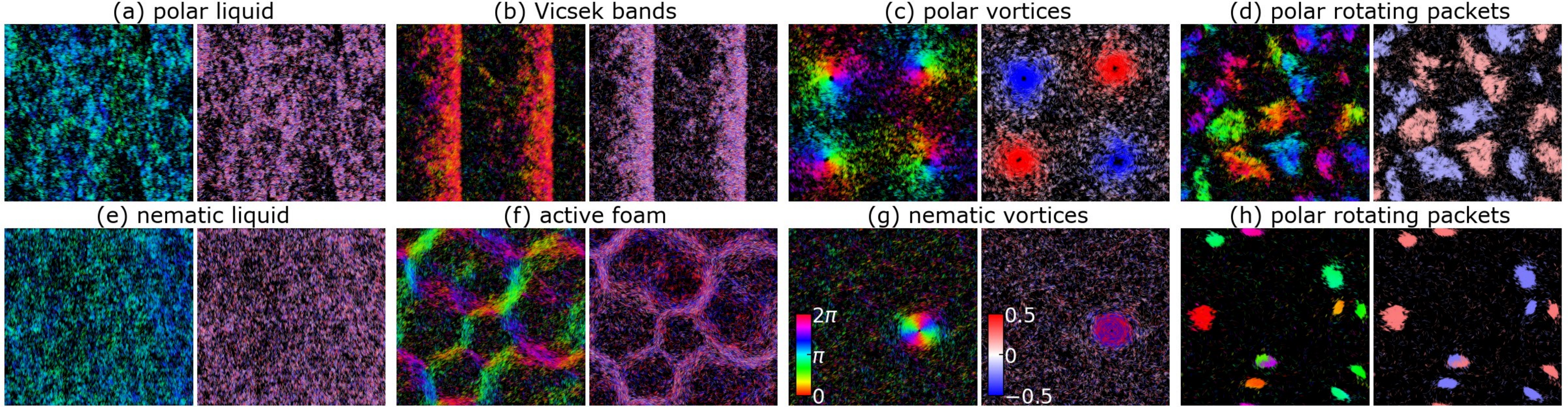}
	\caption{Typical snapshots of phases reported in Fig.~\ref{fig1} taken after transients 
	following random initial conditions.
	These images were obtained in the Gaussian case, except the polar rotating packets (panels (d,h)), 
	which are only observed in the bimodal case.
	For each phase, two subpanels are shown: particles' orientation (left) and intrinsic chirality (right) 
	are represented as small colored segments
	(colormaps are in panel (g)). For (e,f,g), which have local nematic order, $2\theta$ is represented in the left subpanel. For the other phases, the polar angle $\theta$ is shown.
	In the right subpanels, chirality-segregated structures appear red (CW) or blue (CCW), %%%
	while mixed chirality regions appear purple.
	The chirality colormap goes from $-\omega_{\rm max}$ to $+\omega_{\rm max}$, 
	with $\omega_{\rm max}$ adapted for each panel for better legibility.
	(a) polar liquid ($\omega_0=0.004$, $D_{\rm r}=0.03$, $\omega_{\rm max}=0.005$).
	(b) polar (Vicsek) bands ($\omega_0=0.004$, $D_{\rm r}=0.06$, $\omega_{\rm max}=0.005$).
	(c) polar vortices ($\omega_0=0.054$, $D_{\rm r}=0.03$, $\omega_{\rm max}=0.05$).
	(d) rotating polar packets ($\omega_0=0.167$, $D_{\rm r}=0.03$, $\omega_{\rm max}=0.5$).
	(e) nematic liquid ($\omega_0=0.004$, $D_{\rm r}=0.008$, $\omega_{\rm max}=0.005$).
	(f) active foam ($\omega_0=0.03$, $D_{\rm r}=0.016$, $\omega_{\rm max}=0.02$).
	(g) nematic vortices ($\omega_0=0.118$, $D_{\rm r}=0.012$, $\omega_{\rm max}=0.05$).
	(h) rotating polar packets ($\omega_0=0.05$, $D_{\rm r}=0.002$, $\omega_{\rm max}=0.1$).
	}
	\label{fig2}
\end{figure*}
%%%%%%%%%%%%%%%%%%%%%%%%%%%

Chirality disorder introduces another important parameter, the distribution width $\omega_0$. Below we present results obtained at fixed density of particles $\rho_0=1$,  varying $D_{\rm r}$ (with fixed $\kappa=1$) 
and $\omega_0$. 
All simulations presented here were performed at $v_0=1$ using an explicit Euler scheme with timestep $0.1$.
Phase diagrams at fixed, finite size $L=256$ in this $(\omega_0, D_{\rm r})$ plane are presented in Fig.~\ref{fig1} for both 
ferromagnetic ($\alpha=1$) and nematic ($\alpha=2$) alignment, and for both Gaussian and bimodal distribution of
frequencies. Typical snapshots representing most involved phases are shown in Fig.~\ref{fig2}. 
Since here we are chiefly concerned with the fate of the phases present in the pure case under the influence of chirality disorder, a detailed study of the phase diagrams is beyond the scope of this paper 
and will be presented elsewhere \cite{TBP-KVM}. The next paragraph only provides a brief synthetic description,
stressing the similarities between the four cases studied.

The homogeneous ordered liquid phases and the coexistence band phases are broken at strong enough
disorder, giving way to density-segregated phases unknown in the pure case.
With ferromagnetic alignment (Fig.~\ref{fig1}, left), particles are then also spontaneously chirality-segregated into dense, locally polarly ordered CW and CCW structures. In the Gaussian case, only axisymmetric vortices appear
(Fig.~\ref{fig2}(c)) \footnote{These
vortices, not reported in \cite{levis2019activity,levis2019simultaneous}, are somewhat reminiscent of the `chiral clusters' found in \cite{moore2021chiral}.}. 
In the bimodal case, vortices themselves give way, at higher $\omega_0$ values, 
to rotating polar packets (Fig.~\ref{fig2}(d)), which have a global polarity that rotates in time (contrary to vortices).
With nematic alignment (Fig.~\ref{fig1}, right), 
similarly to the ferromagnetic case, chirality-sorted rotating polar packets are observed with bimodal disorder
(Fig.~\ref{fig2}(h)). 
Other phases are not chirality-sorted. 
For both types of disorder, well-formed nematic vortices (Fig.~\ref{fig2}(g)) are present 
for $D_{\rm r}$ values roughly corresponding to the range over which the homogeneous nematic 
is observed in the pure case. For stronger noise, the nematic bands of the pure case become an active foam, 
i.e.  a constantly-rearranging network of thin nematic bands (Fig.~\ref{fig2}(f)).

\begin{figure}[b!]
\includegraphics[width=\columnwidth]{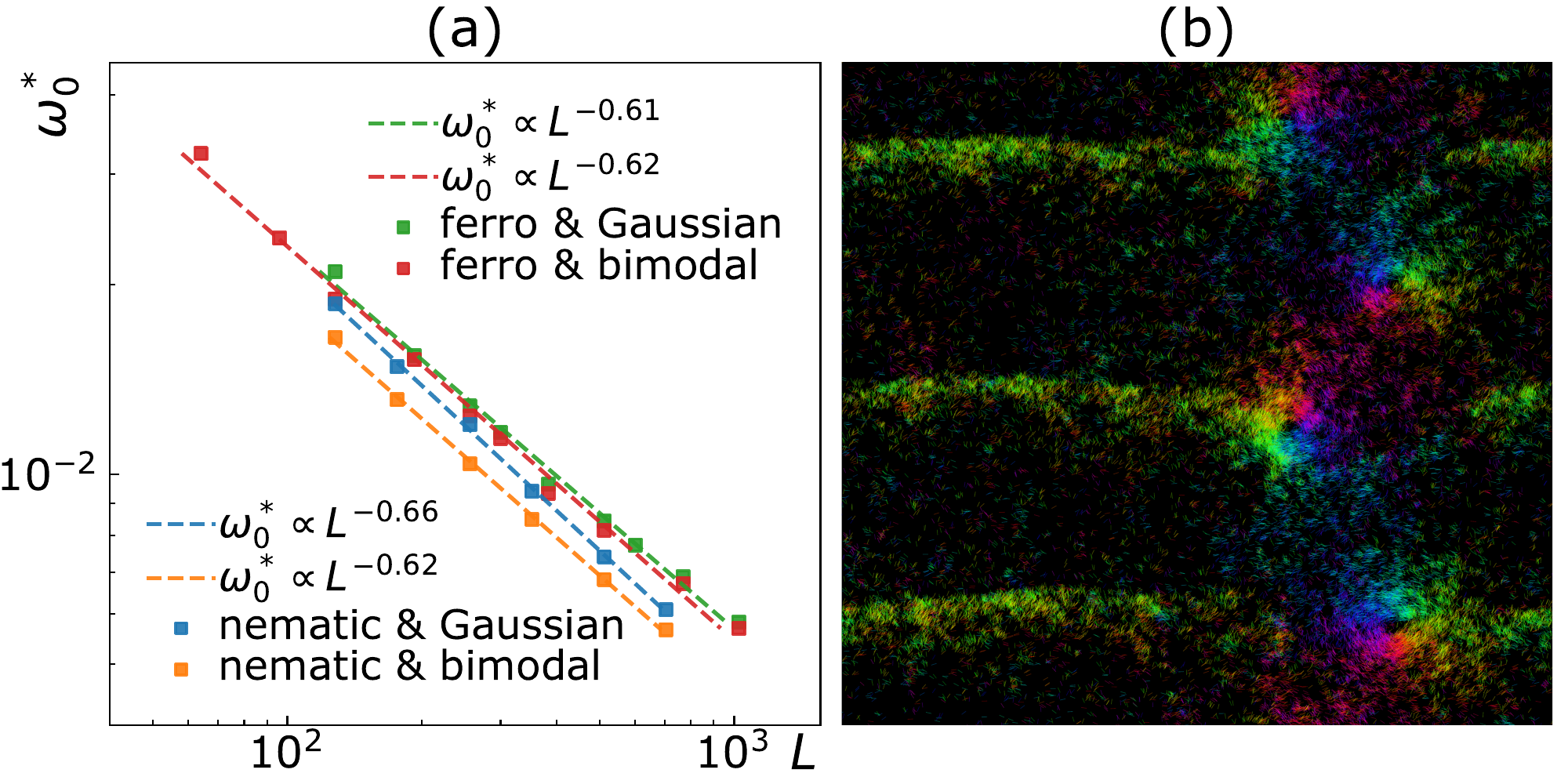}
	\caption{(a) Variation of the $[0,\omega_0^*(L)]$ domain of existence of the homogeneous ordered phases with system size $L$ (ferromagnetic cases $D_{\rm r}=0.03$, nematic cases $D_{\rm r}=0.004$). 
	(b) snapshot taken in the region of coexistence between Vicsek bands and polar vortices 
	(ferromagnetic alignment, Gaussian distribution of chiralities, $\omega_0=0.06$, 
	$D_{\rm r}=0.054$, $L=256$, colors indicate particles' orientation as in Fig.~\ref{fig2}(a-d) left subpanels).
	}
	\label{fig3}
\end{figure}

As seen above for $L=256$, 
the pure-case ordered liquid phases are observable at finite values of $\omega_0$ 
(examples are given in Fig.~\ref{fig2}(a,e)). The presence of chirality disorder is reflected
in the bimodal distribution of the orientations of particles (not shown, but see \cite{levis2019activity}).
However, the liquid phases do not survive the $L\to\infty$ limit:
choosing $D_{\rm r}$ values well into the liquid phase in the pure case, we estimated, for various system sizes,
the maximum disorder value $\omega_0^*$ beyond which global order breaks down due to 
the emergence of vortices~\footnote{Increasing $\omega_0$, the time-averaged  
polar or nematic order parameter typically decays rather suddenly from $O(1)$ to vanishing values 
when the vortex phase is reached. The threshold $\omega_0^*$ is then estimated to be at the middle point of
this sudden decay.}
As shown in Fig.~\ref{fig3}(a), 
$\omega_0^* \sim L^{-\gamma}$ with $\gamma\simeq 0.6$, indicating that the parameter region where the
ordered fluid can be observed shrinks with increasing system size, both in the ferromagnetic and nematic cases, 
for both Gaussian and bimodal disorder.

In contrast, $\omega_0^*$ does not vanish when $L\to\infty$ at a $D_{\rm r}$ value chosen
to be at the level of the traveling (Vicsek) band phase (not shown). 
For large systems with ferromagnetic alignment, 
one observes the band-vortices transition at more or less the same $\omega_0^*$ value, 
but with a region of coexistence between the two phases (Fig.~\ref{fig3}(b)). The Vicsek bands are thus
robust structures that can survive chirality disorder and even coexist with chiral structures.

To be complete, let us describe the fate of the nematic coexistence phase. 
In systems of moderate size, one typically observes a single nematic band in the pure case, but it is well known that
nematic bands are inherently unstable, with this instability leading to band spatiotemporal chaos \cite{ngo2014large,grossmann2016mesoscale,cai2019dynamical}. 
Thus the nematic coexistence phase is intrinsically disordered even in the pure case. 
Chirality disorder, as indicated in our phase diagrams (Fig.~\ref{fig1}), leads to the dynamic active foam 
illustrated in Fig.~\ref{fig2}(f). This regime appears to be different from band chaos, 
and is reminiscent of the active foams described in \cite{nagai2015collective,maryshev2020pattern}
but this point will be studied elsewhere \cite{TBP-KVM}

A major conclusion of the numerical study above is that the homogeneous ordered phases 
(polar flocks and nematic) seem to be broken, asymptotically,
by any amount of chirality disorder. 
We now turn to a theoretical understanding of this at the continuous level. 
We focus on the case of ferromagnetic alignment and bimodal distribution of chiralities 
for which it is easiest to construct a hydrodynamic theory starting from our microscopic model \eqref{eq:kvm}.
We follow the Boltzmann-Ginzburg-Landau approach \cite{bertin2013mesoscopic,peshkov2014boltzmann,DADAM_LesHouches,chate2020dry}.
Details are given in \cite{SUPP}.

%%%%%%%%%%%%%%%%%%%%%%%%%%%
%%%%%%%%%%%%%%%%%%%%%%%%%%%

In the bimodal case, it is quite natural to separate the $+\omega_0$ and $-\omega_0$ subpopulations.
We write 2 coupled Boltzmann equations ruling the 
evolution of their one-body probability density functions $f^+({\bf r},\theta,t)$ and $f^-({\bf r},\theta,t)$:
\begin{equation} \label{eq_Boltzmann}
\partial_t f^\pm + v_0{\bf e}(\theta)\!\cdot\!\nabla f^\pm \pm\omega_0 \partial_\theta f^\pm = 
 I_{\rm sd}[f^\pm] + I_{\rm co}[f^\pm\!,f],
\end{equation}
where $f=f^+ + f^-$, and $I_{\rm sd}$ and $I_{\rm co}$ are self-diffusion and 
collision integrals given in \cite{SUPP}. 
%%%
Note that the two equations are only coupled via the collision integral, and each of them is thus similar to the
Boltzmann equation for the pure case \cite{peshkov2014boltzmann}.
Expanding $f^\pm$ in Fourier series of $\theta$ 
(i.e. $f^\pm({\bf r},\theta,t)=\tfrac{1}{2\pi}\sum_{k=-\infty}^{+\infty} f^\pm_k({\bf r},t)e^{-ik\theta}$)
the Boltzmann equations are de-dimensionalized and transformed into a hierarchy of partial differential equations 
for the complex fields $f^+_k$ and $f^-_k$. 
Linear stability analysis of the disordered solution 
$\rho^\pm\equiv f^\pm_0=\tfrac{1}{2}\rho_0$, $f^\pm_{k\ne0}=0$
reveals that it is unstable to $f^\pm_1$ perturbations at large density and/or weak noise. 
Thus, unsurprisingly, local polar order emerges at onset and one can truncate and close
the hierarchy of equations using the same Ginzburg-Landau scaling ansatz as in the pure case.
Denoting, for legibility, $p\equiv f^+_1$ and $m \equiv f^-_1$, we obtain:
\begin{subequations}
\label{eq:hydro}
\begin{align}
\partial_t \rho^+ =& - {\rm Re}[ \triangledown^* p] \,, \label{eq:rho}\\ 
\partial_t p  = &  \left( \mu[\rho^+ ,\rho^-] - i \omega_0 - \xi |p|^2 \right) p + \nu \Delta p  \nonumber \\
+& \kappa_1 \triangledown^* p^2 + \kappa_2 p^* \triangledown p -\tfrac{1}{2} \triangledown \rho^+\nonumber \\
+& ( \tilde{\mu}[\rho^+]- \tilde{\xi} |m|^2 ) m + \tilde{\nu} \Delta m  
+ \tilde{\kappa}_1 \triangledown^* \! m^2 + \tilde{\kappa}_2 m^*\! \triangledown m \nonumber \\
+& \gamma_1 |m|^2 p + \gamma_2 m^2 p^*
+ \tilde{\gamma}_1 |p|^2 m + \tilde{\gamma}_2 p^2 m^* \nonumber \\
+& \delta_1 \triangledown^*(pm) +  \delta_2 p^* \triangledown m + \tilde{\delta}_2 m^* \triangledown p
\,, \label{eq:p}
\end{align}
\end{subequations}
where $\triangledown \equiv \partial_x + i\partial_y$ denotes the complex gradient, 
$\Delta=\triangledown\triangledown^*$ is the Laplacian in this complex notation,
and the dependence of coefficients on local density have been explicited. 
(The equations for $\rho^-$ and $m$ are obtained by performing the swaps $\rho^+ \leftrightarrow \rho^-$,
$p \leftrightarrow m$, and $\omega_0 \leftrightarrow -\omega_0$.) 
A few comments are in order: \eqref{eq:rho} {\it is} the usual exact conservation equation 
($\partial_t\rho+\nabla(\rho{\bf v})=0$). 
In contrast with the Toner-Tu equation of the pure case, 
almost all coefficients in \eqref{eq:p} are complex, as shown in \cite{SUPP} where their expression
in terms of the microscopic parameters is given. The first two lines of \eqref{eq:p} are the Toner-Tu equation of the pure case (up to the $i\omega_0$ term and the complex nature of some coefficients); 
the next line has similar terms but for the $m$ field, 
while the last two lines regroup the terms coupling $p$ and $m$. 

\begin{figure}
\includegraphics[width=\columnwidth]{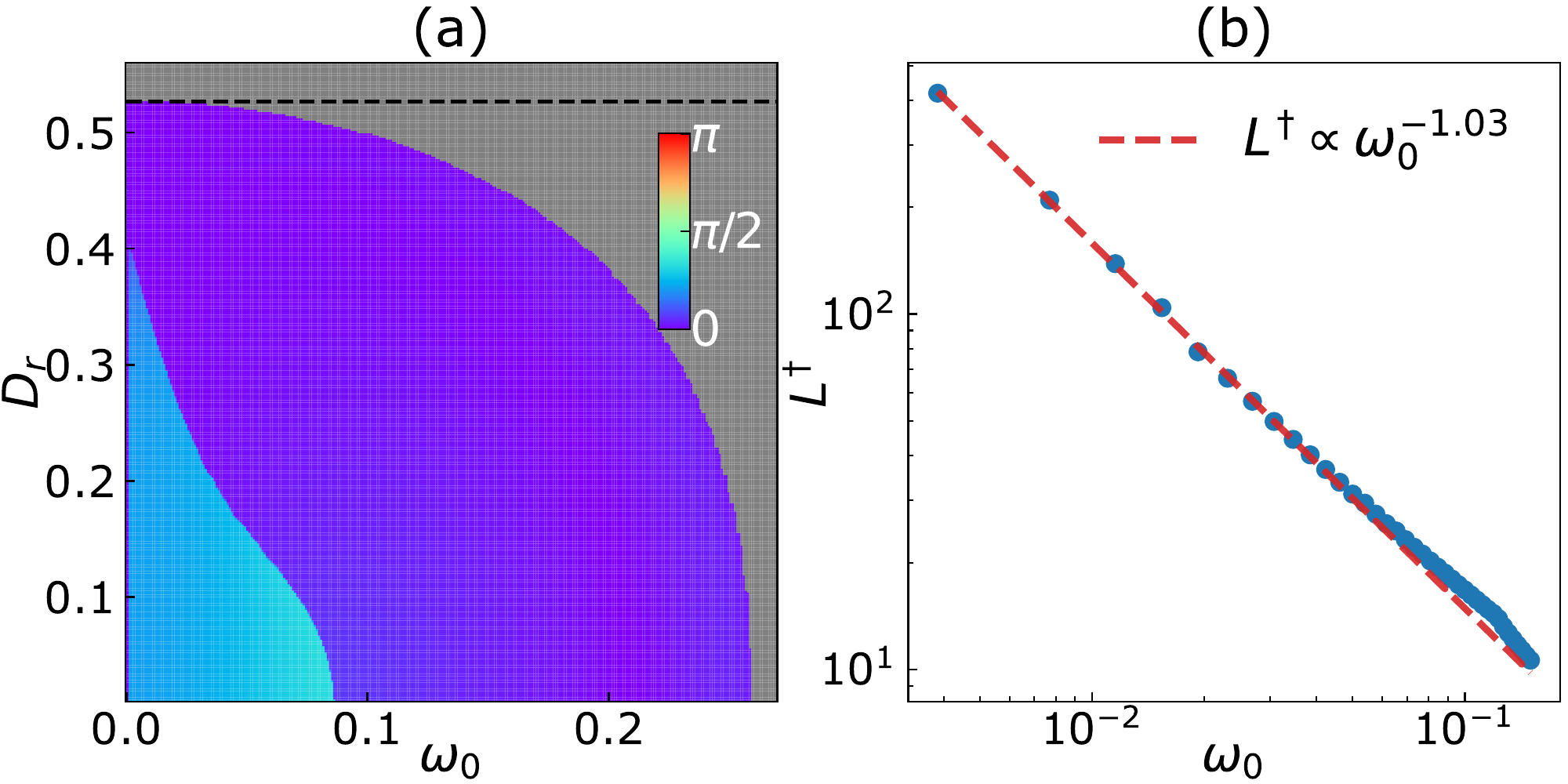}
	\caption{Linear stability analysis of the homogeneous ordered solution of the hydrodynamic theory
	($\rho^+=\rho^-=1$).
	(a): direction of most unstable mode in $(\omega_0, D_r)$ plane (in the grey area, the solution does not exist;
	the homogeneous disordered solution is linearly unstable below the dashed line). 
	(b): variation of $L^\dag$ (size of largest stable system) with $\omega_0$ at fixed $D_{\rm r}=0.4$.
	}
	\label{fig4}
\end{figure}

Below, we only study the linear stability of the stationary homogeneous solutions of the above equations.
The comprehensive numerical study at the nonlinear, inhomogeneous level is ongoing work that will appear elsewhere
\cite{TBP-KVM}.
Apart from the trivial disordered solution $\rho^\pm=\tfrac{\rho_0}{2}, p=m=0$, Eqs.~(\ref{eq:rho},\ref{eq:p}) 
and their counterparts governing $\rho^-$ and $m$ have another homogeneous solution with
$|p|=|m|=P, m/p=\exp(i\Omega)$. This ordered solution, 
described in \cite{levis2019activity} in some limit case of the above hydrodynamic equations, reduces to the
polar flock solution of the Toner-Tu equations  in the $\omega_0\to 0$ limit where $\Omega=0$. 
It corresponds to the polar flock phase observed at particle level (Fig.~\ref{fig1}(c)). 
Its expression in closed form is cumbersome, and in practice we calculate it numerically at arbitrary precision 
from the simple equations defining $P$ and $\Omega$. 
This allows to determine not only its existence domain, but also its full linear stability analysis.
While all details are provided in \cite{SUPP}, we only sketch here how this is done.
We first linearize the equations around the ordered solution and write the resulting $6\times6$
matrix in Fourier space, where its coefficients depend on a wavenumber ${\bm q}=(q_\|, q_\perp)$, 
here expressed in the coordinates relative to the order of the solution
\footnote{We do not seek to simplify the matrix by taking the small $q=|{\bm q}|$ limit and/or specializing to longitudinal or transversal directions, as this may be misleading.}.
For each parameter set of interest, we solve the matrix and determine the most unstable (or least stable) mode
${\bm q}^*$, i.e. the mode with the highest growth rate $\sigma^*$. 
We find that the solution is unstable ($\sigma^*>0$) {\it everywhere} in its domain of existence
(Fig.~\ref{fig4}(a)). 
%%%
This indicates that the instability of the polar flock solution found at particle level (Fig.~\ref{fig3}(a)) 
is probably due to this linear instability at (deterministic) hydrodynamic level. 

We also determined the wavevectors separating stable from unstable modes, and in particular that with the largest
wavenumber $q^\dag=|{\bm q}^\dag|$, which is of interest since $L^\dag=1/q^\dag$ is the maximum system size 
at which the ordered solution is stable. 
We find that $L^\dag$ varies like $1/\omega_0$ (Fig.~\ref{fig4}(b)).
We believe this algebraic law is at the root of the one observed at microscopic level
(Fig.~\ref{fig3}(a)). The fact that the scaling exponent takes a 'non-trivial' value in this last case
maybe due to fluctuations and nonlinear effects.

%%%%%%%%%%%%%%%%%%%%%%%%%%%
%%%%%%%%%%%%%%%%%%%%%%%%%%%

To summarize, we have shown that the homogeneous orientationally-ordered phases of dry aligning active matter,
be they polarly or nematically ordered,
are susceptible to any amount of population heterogeneity introduced in the form of chirality disorder. 
Even though finite systems may resist a finite amount of such disorder, 
the maximum disorder strength that can be supported by the ordered liquid vanishes
in the infinite-size limit. 
We have traced this back, in the polar case, to the generic instability of the ordered homogeneous solution
of the hydrodynamic theory that we derived from our particle-level model.
Our results (in the polar, ferromagnetic case) contradict the ``activity-induced synchronisation" put forward in \cite{levis2019activity}. This paper, though, did not present a finite-size study such as that shown in Fig.~\ref{fig3},
nor a linear stability analysis of the ordered solution as in Fig.~\ref{fig4}
\footnote{To be true, Ref.~\cite{levis2019activity} deals with a version of Eqs.~\eqref{eq:kvm}
where the interaction term is {\it not} normalized and presents a hydrodynamic theory obtained 
via a ``Fokker-Planck'' approach that is slightly different (less terms) than our Eqs.~\eqref{eq:hydro}.
We have checked that within this simpler hydrodynamic theory the ordered homogeneous solution is prone to the same generic instability as the one found with Eqs.~\eqref{eq:hydro}. Details will appear in \cite{TBP-KVM}.}.

Thus, as in variants of the Kuramoto model where random oscillators are locally coupled, 
synchronization of random frequency/chirality active particles is impossible in 2D, 
in spite of the true long-range order (aka synchronization) 
proven by Toner and Tu in the pure case. 
In this context, the study of higher-dimensional systems
would be interesting since it is known that, for the locally-coupled Kuramoto model,
frequency-synchronization is possible in three and four dimensions, 
whereas phase-synchronisation occurs above \cite{hong2007entrainment}. 
Ongoing work investigates ``flocking'' versions of the models of globally-cioupled high-dimension oscillators
studied in \cite{chandra2019continuous,zheng2021transition}.

\acknowledgments
We thank Yu Duan, Beno\^{\i}t Mahault, Alexandre Solon, and Yongfeng Zhao 
for a critical reading of an earlier version of this paper.
We acknowledge generous allocations of cpu time on Beijing CSRC’s Tianhe supercomputer.
This work is supported by the National Natural Science Foundation of China 
(Grants \# 11635002 to X.-q.S. and H.C., 11922506 and 11674236 to X.-q.S.)

\bibliography{./Biblio-current.bib}

\end{document}